\documentclass[aps,11pt]{revtex4}
\usepackage[dvips]{graphicx}
\usepackage{amsmath}
\usepackage{bm}
\usepackage{times}
\usepackage{epsfig}

\def\beq{\begin{equation}}
\def\eeq{\end{equation}}
\def\bea{\begin{eqnarray}}
\def\eea{\end{eqnarray}}

\def\nn{\nonumber}
\def\Eq#1{Eq.~(\ref{#1})}

\begin{document}

\preprint{IFIC/08-18}
\vspace{2.0cm}
\title{\Large {\textit{Proton Stability, Dark Matter and Light Color
Octet Scalars \\ in Adjoint SU(5) Unification}}}
\vspace{4.0cm}
\author{Pavel Fileviez P\'erez}
\affiliation{University of Wisconsin-Madison, Department of Physics\\
1150 University Avenue, Madison, WI 53706, U.S.A.}
\email{fileviez@physics.wisc.edu}
\author{Hoernisa Iminniyaz}
\affiliation{Physics Department, University of Xinjiang,
830046 Urumqi, P.R. China.}
\email{hoernisa@th.physik.uni-bonn.de}
\author{Germ\'an Rodrigo}
\affiliation{Instituto de F\'{\i}sica Corpuscular, CSIC-Universitat de
Val\`encia, Apdo. de Correos 22085, E-46071 Valencia, Spain.}
\email{german.rodrigo@ific.uv.es}

\date{\today}
\begin{abstract}
\textit{The unification of gauge interactions in the context of
Adjoint SU(5) and its phenomenological consequences are investigated.
We show the allowed mass spectrum 
of the theory which is compatible with proton 
decay, and discuss the possibility to have a cold dark matter candidate. 
Due to the upper bounds on the proton decay partial lifetimes,
$\tau( p \to K^+ \bar{\nu}) \ \leq 9.3 \times 10^{36}\ $ years and
$\tau( p \to \pi^+ \bar{\nu}) \ \leq \ 3.0 \times 10^{35}$ years,
the theory could be tested at future proton decay experiments.
The theory predicts also light scalar color octets which could be
produced at the Large Hadron Collider.}
\end{abstract}
\maketitle

\section{Introduction}
The possibility to have the unification of the gauge interactions
has been a guiding principle for physics beyond the Standard Model (SM)
since the paper by Georgi, Quinn and Weinberg in 1974~\cite{GW}.
The simplest grand unified theory (GUT) was proposed the same
year in a seminal paper~\cite{GG}. As it is well known, this theory
is based on the $SU(5)$ gauge symmetry and the SM
matter of one family is unified in two different representations.
Unfortunately, this simple GUT theory is ruled out by the
present values of the coupling constants. At the same time, in
the original version of the theory~\cite{GG} there are no
massive neutrinos and a relation between the Yukawa couplings
for the charged leptons and down quarks, $Y_E=Y_D^T$, is predicted,
which is in disagreement with the experiments.

One of the main predictions of GUT theories is the decay of the
lightest baryon~\cite{PS}. Thanks to the great effort of many
experimental collaborations one has today very impressive lower bounds
on the different partial decay lifetimes~\cite{PDG}.
In the future, the searches for proton decay will continue and hopefully
it will be finally discovered. See for example~\cite{Experiments} for
a review of future experiments. From the theoretical point of view
it is very difficult to make strong predictions for the lifetime of the proton
since in most of the theories the grand unified scale needs to be predicted
with great precision, the nucleon-meson matrix elements are not well-known,
or the supersymmetric character of the theory might led to a large model
dependence. See~\cite{review} for a review of proton decay predictions
in several theoretical frameworks. Since proton decay is a smoking
gun signature for GUT theories one should try to make the best
predictions in the simplest scenarios.

As we have explained above the simplest, but not realistic,
GUT theory was proposed by Georgi and Glashow in Ref.~\cite{GG}.
If one sticks to renormalizability there are, however,
three possible ways out to construct realistic theories:
\begin{itemize}
\item \textit{Renormalizable $SU(5)$ with Type I seesaw}:
As it has been known for a long time in order to have a consistent
model for charged fermion masses at the renormalizable
level one has to introduce an extra Higgs in the $\bm{45}$
representation~\cite{GJ}. At the same time, at least two fermionic
singlets are needed to generate neutrino
masses through the Type I seesaw mechanism~\cite{TypeI}.
This scenario has been studied in detail in~\cite{Ksbabu}
and~\cite{Ilja-Pavel-45}. From the results presented
in~\cite{Ilja-Pavel-45} it is possible to conclude that
once one imposes the most conservative bound on the
mass of the Higgses mediating proton decay this theory
is ruled out by proton decay experiments.

\item \textit{Renormalizable $SU(5)$ with Type II seesaw}:
In this scenario the Higgs sector is composed of $\bm{5}_H$,
$\bm{15}_H$, $\bm{24}_H$ and $\bm{45}_H$,
and  the neutrino masses are generated
through the Type II seesaw mechanism~\cite{TypeII}. One can conclude already
that this model is not appealing since the Higgs sector is very
complicated. See~\cite{Ilja-Pavel-45} for a recent discussion
of this scenario.

\item \textit{{Renormalizable Adjoint $SU(5)$}}:
This is the simplest realistic renormalizable theory based
on $SU(5)$~\cite{adjoint}.
In this scenario the Higgs sector is composed
of $\bm{5}_H$, $\bm{24}_H$ and $\bm{45}_H$ and an extra
fermionic representation in the $\bm{24}$ adjoint representation
is introduced in order to generate the neutrino masses
through the Type I~\cite{TypeI} and Type III~\cite{TypeIII}
seesaw mechanisms. In this theory, one has a new realization
of the Type III seesaw mechanism in the context
of a renormalizable GUT theory where
only two Higgses, $\bm{5}_H$ and $\bm{45}_H$,
generate all fermion masses. For the supersymmetric
version of the theory see~\cite{SUSY-adjoint}.

\end{itemize}
Since the theory proposed in Ref.~\cite{adjoint} is the simplest
renormalizable $SU(5)$ theory, we study its possible phenomenological
predictions in detail. We study for the first time the properties of
the full Lagrangian of the theory, computing all fermion masses.
We find that there is only one possible scenario for the masses
of the fermions in the adjoint representation which is allowed
by unification and proton decay. We notice that the neutral
component of the real scalar triplet living in the adjoint
representation could be a possible candidate for the Cold Dark
Matter (CDM) in the Universe. Using the proton decay
and dark matter constraints we find that the possible solutions
for the mass spectrum in the theory allowed by unification
are reduced considerably. Using these results we
predict upper bounds on the partial proton decay lifetimes,
that can be tested at future proton decay experiments.
The possibility to have light exotic fields,
like scalar color octets, is discussed.

The outline of the paper is as follows: In Section II we discuss
the main properties of {\textit{Adjoint}} $SU(5)$ and its main
predictions. The predictions coming from the unification of gauge
interactions are discussed in Section III. The dark matter constraints
and the upper bounds on the proton decay partial lifetimes are
discussed in Sections IV and V, respectively. The possible
light exotic fields are pointed out in Section VI, while in
Section VII we summarize our results. In the Appendix A we
write down all details of the theory, the matter content
and all interactions. Finally, in Appendix B we list all
contributions of the fields present in the theory
to the running of gauge couplings.

\section{Adjoint $\bm{SU(5)}$}
A new renormalizable GUT theory based on the $SU(5)$ symmetry
where the neutrino masses are generated through the Type I and Type III
seesaw mechanisms has been recently proposed in Ref.~\cite{adjoint}.
In this section we discuss the main properties
of this theory in order to understand its main phenomenological
predictions.
\begin{itemize}
\item \underline{Matter Unification}: As in any theory based on $SU(5)$,
the SM matter is unified in
the ${\bm {\overline 5}} = l_L \bigoplus (d^C)_L$ and
${\bm {10}} = (u^C)_L \bigoplus q_L \bigoplus (e^C)_L$ representations.
In order to generate the neutrino masses through the Type I~\cite{TypeI} and
Type III~\cite{TypeIII} seesaw mechanisms an extra matter
field in the adjoint representation
${\bm{24}} = (\rho_8)_L \bigoplus (\rho_3)_L \bigoplus (\rho_{(3,2)})_L
\bigoplus (\rho_{(\bar{3}, 2)})_L \bigoplus (\rho_{0})_L$
is introduced. Notice that since the
extra matter is in the adjoint representation it does not induce anomalies,
making the model very simple and appealing. Here, it is important to mention
that the field $\rho_8$ must be heavier than $10^6-10^7$ GeV in order to
satisfy the constraints coming from Big Bang Nucleosynthesis, once
one assumes that the unification scale is larger than $3 \times 10^{15}$ GeV,
which is consistent with the experimental lower bounds on the
proton decay lifetime.

\item \underline{Higgs Sector}: The Higgs sector
is composed of the representations $\bm{5}_H = H_1 \bigoplus T$,
$\bm{24}_H  =  \Sigma_8 \bigoplus \Sigma_3 \bigoplus \Sigma_{(3,2)}
\bigoplus \Sigma_{(\overline{3},2)} \bigoplus \Sigma_{24}$,
and
$\bm{45}_H = \Phi_1 \bigoplus \Phi_2 \bigoplus \Phi_3 \bigoplus
\Phi_4 \bigoplus \Phi_5 \bigoplus \Phi_6 \bigoplus H_2$.
As it is well-known, this is the minimal Higgs sector of any
renormalizable model compatible with the spectrum of the charged fermions.
The main difference with respect to other theories
is that once the $\bm{24}$ representation is introduced the
$\bm{45}_H$ field plays a crucial role to generate
neutrino masses. See Appendix A for all interactions in the Higgs sector.

\item \underline{SM Fermion Masses}: In this theory all the fermion masses
are generated at the renormalizable level. The relevant Yukawa
interactions are given by
\begin{eqnarray}
- S_{\rm Yukawa} &=&
\int d^4 x
\Bigg( Y_1 \ 10 ~\bar{5} ~5^*_H
\ + \
Y_2 10 ~\bar{5} ~45^*_H
\ + \ Y_3 \ 10~ 10 ~ 5_H  \
+ \
Y_4 \ 10 ~10 ~45_H \Bigg) \ + \
\nonumber \\
&+&  \int d^4 x \Bigg( c \ \bar{5} \ 24 \ 5_H \ + \
p \ \bar{5} \ 24 \ 45_H \Bigg) \ + \ \text{h.c.}
\label{Yukawa}
\end{eqnarray}
The masses of the SM charged fermions read
\begin{eqnarray}
M_D & = & Y_1 \frac{v_5^*}{\sqrt{2}} \ + \ 2 \ Y_2 \frac{v_{45}^*}{\sqrt{2}}~, \\
M_E & = & Y_1^T \frac{v_5^*}{\sqrt{2}} \ - \ 6 \ Y_2^T \frac{v_{45}^*}{\sqrt{2}}~, \\
M_U & = & 4 \left( Y_3 + Y_3^T \right) \frac{v_5}{\sqrt{2}} \ - \ 8 \ \left( Y_4 - Y_4^T \right)
\frac{v_{45}}{\sqrt{2}}~,
\end{eqnarray}
with $v_5$ and $v_{45}$ being the vacuum expectation values
of $\bm{5}_H$ and $\bm{45}_H$, respectively.
See Appendix A for details. The neutrino mass matrix is
\begin{eqnarray}
M^\nu_{ij} & = & \frac{a_i a_j}{M_{\rho_3}} \ + \ \frac{b_i
 b_j}{M_{\rho_0}}~,
\label{neutrinos}
\end{eqnarray}
where
\begin{equation}
a_i = \frac{1}{2\sqrt{2}} \left( c_i  v_5 \ - \  3 p_i v_{45} \right) \qquad \text{and} \qquad
b_i  =  \frac{\sqrt{15}}{2 \sqrt{2}} \left( \frac{c_i  v_5}{5} \ + \ p_i v_{45}
  \right)~.
\label{n-couplings}
\end{equation}
One of the neutrinos is massless. The additional interactions
\begin{eqnarray}
S_{24} &=&  - \int d^4 x \Bigg( {M} \ \text{Tr} \ 24^2
\ + \ {\lambda} \ \text{Tr} \left( 24^2 \ 24_H \right) \Bigg) \ + \ \text{h.c.}
\label{seesaw}
\end{eqnarray}
give mass to the fermions living in the adjoint representation,
and in particular to $\rho_0$ and $\rho_3$, the fields responsible
for the seesaw in \Eq{neutrinos}. Their masses are given by
\bea
M_{\rho_0} &=& \left| \ m \ - \ e^{i \alpha} \ \Lambda \right|~,
\qquad
M_{\rho_3} = \left| \ m \ - \ 3 \ e^{i \alpha} \ \Lambda \right|~,
\nn
\\
M_{\rho_8} &=& \left| \ m \ + \ 2 \ e^{i \alpha} \ \Lambda \right|~,
\qquad \text{and} \qquad
M_{\rho_{(3,2)}} = M_{\rho_{(\bar{3},2)}} = \left| \ m \ -
\ \frac{1}{2} \ e^{i \alpha} \ \Lambda \right|~.
\label{masses}
\eea
Here we have used the relations $M_V= v \sqrt{5 \pi \alpha_{GUT}/3}$ and
$\Lambda=\tilde{\lambda} \, M_{GUT}/\sqrt{\alpha_{GUT}}$ with
$\tilde{\lambda}= |\lambda| / {\sqrt{50 \pi}}$, and have chosen the mass
of the superheavy gauge boson $M_V$ as the unification scale.
The phase $\alpha$ is the relative phase between $M$ and the
coupling $\lambda$, while $m=|M|$.

\item \underline{Proton Decay}:
There are several fields contributing to proton decay.
The dimension six gauge contributions are
mediated by the superheavy gauge bosons
$V \sim (\bm{3},\bm{2},-5/6)\bigoplus(\overline{\bm{3}},\bm{2},5/6)$,
which must be heavier than $3 \times 10^{15}$ GeV in order to satisfy the
experimental lower bound on the proton decay lifetime.
The $SU(3)$ triplets $T = (\bm{3},\bm{1},-1/3)$,
$\Phi_3 = (\bm{3},\bm{3},-1/3)$,
$\Phi_5 = (\bm{3},\bm{1}, -1/3)$ and
$\Phi_6 = (\overline{\bm{3}}, \bm{1},4/3)$
mediate the dimension six Higgs contributions.
The most conservative lower bound on their masses
from proton decay is
$M_{T}, M_{\Phi_3}, M_{\Phi_5}, M_{\Phi_6} > 10^{12}$ GeV.
These bounds are very important in order to understand the possible
solutions for the spectrum which are allowed by unification
and proton decay.
\end{itemize}

\section{UNIFICATION OF GAUGE INTERACTIONS}
In this section we investigate the main phenomenological predictions
due to the unification of gauge couplings.
The contribution of all the fields to the
running of the gauge couplings are listed in Table I of Appendix B.
The fields $\rho_3$, $\Sigma_3$ and $\Phi_3$ can favor unification
because they have negative (positive) contribution to
$b_1 - b_2 (b_2-b_3)$. Here, $b_i$ stands for the different
beta functions. The field $\Phi_1$ plays also an important role
since it has a negative contribution to $b_1 - b_2$, and helps to increase
the GUT scale such that it might become compatible with proton decay.
In order to simplify our analysis we set all the fields listed in
Table I that do not help to achieve unification at the GUT scale,
and keep only $\Sigma_3$, $\Phi_1$, $\Phi_3$ and the fermionic fields
in the $\bm{24}$ representation below. Then, we study
the possibility to achieve unification in agreement with experiment,
and evaluate the maximal allowed GUT scale.

The relevant equations for the running
of the gauge couplings at one-loop level are:
\bea
&&
\alpha_1^{-1} (M_Z) = \alpha^{-1}_{GUT}
+ \frac{1}{2\pi} \left( b_1^{SM} \ln \frac{M_{GUT}}{M_Z}
+ \frac{1}{5} \ln \frac{M_{GUT}}{M_{\Phi_3}}
+ \frac{4}{5} \ln \frac{M_{GUT}}{M_{\Phi_1}}
  + \frac{10}{3} \ln \frac{M_{GUT}}{M_{\rho_{(3,2)}}} \right), \nn \\
&&
\alpha_2^{-1} (M_Z) = \alpha^{-1}_{GUT} + \nn
\\
 && \quad
+ \frac{1}{2\pi} \Bigg( b_2^{SM} \ln \frac{M_{GUT}}{M_Z} +
  \frac{4}{3} \ln \frac{M_{GUT}}{M_{\rho_3}}
  + \frac{1}{3} \ln \frac{M_{GUT}}{M_{\Sigma_3}}
  + {2} \ln \frac{M_{GUT}}{M_{\Phi_3}}
+ \frac{4}{3} \ln \frac{M_{GUT}}{M_{\Phi_1}}
  + 2 \ln \frac{M_{GUT}}{M_{\rho_{(3,2)}}} \Bigg)~, \nn
\eea
and
\bea
\alpha_3^{-1} (M_Z) &=& \alpha^{-1}_{GUT} +  \frac{1}{2\pi} \Bigg( b_3^{SM} \ln \frac{M_{GUT}}{M_Z} +
\frac{1}{2} \ln \frac{M_{GUT}}{M_{\Phi_3}}
+ {2} \ln \frac{M_{GUT}}{M_{\Phi_1}}
  + \frac{4}{3} \ln \frac{M_{GUT}}{M_{\rho_{(3,2)}}}
  + 2 \ln \frac{M_{GUT}}{M_{\rho_{8}}} \Bigg)~,\nn \\
\label{RGE}
\eea
where $b_1^{SM}=41/10$, $b_2^{SM}=-19/6$ and $b_3^{SM}=-7$
are the SM beta functions.
The masses of the fields living in the $\bm{24}$ representation
are linked by \Eq{masses}. These relations constrain the
possible solutions for unification. Indeed, due to \Eq{masses}
and because we are dealing  with a complete representation
their contributions to the running of the gauge couplings do
not depend on the absolute value of their masses, but only on
the mass splitting between them. For different values of
$m$, $\lambda$ and $\alpha$ we can achieve different benchmark scenarios.

\begin{figure}[ht]
\begin{center}
\includegraphics[width=9.0cm]{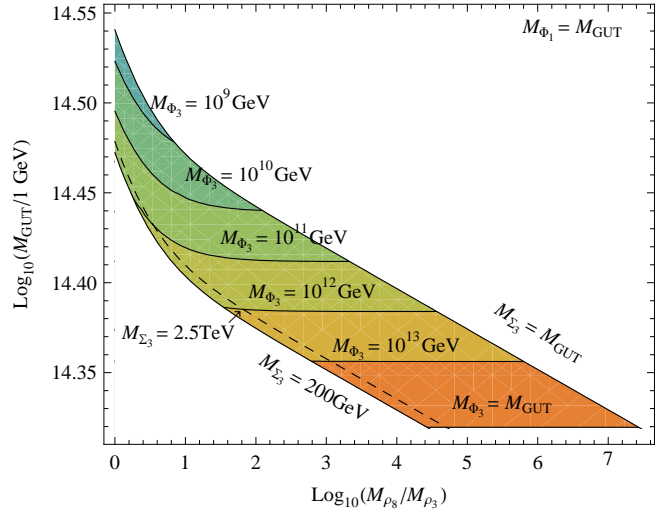}
\caption{Parameter space allowed by unification at the one-loop level
in Adjoint $SU(5)$ for the Benchmark I scenario when $M_{\Phi_1}=M_{GUT}$.}
\label{fig:alpha0GUT}
\end{center}
\end{figure}

\begin{figure}[ht]
\begin{center}
\includegraphics[width=9.0cm]{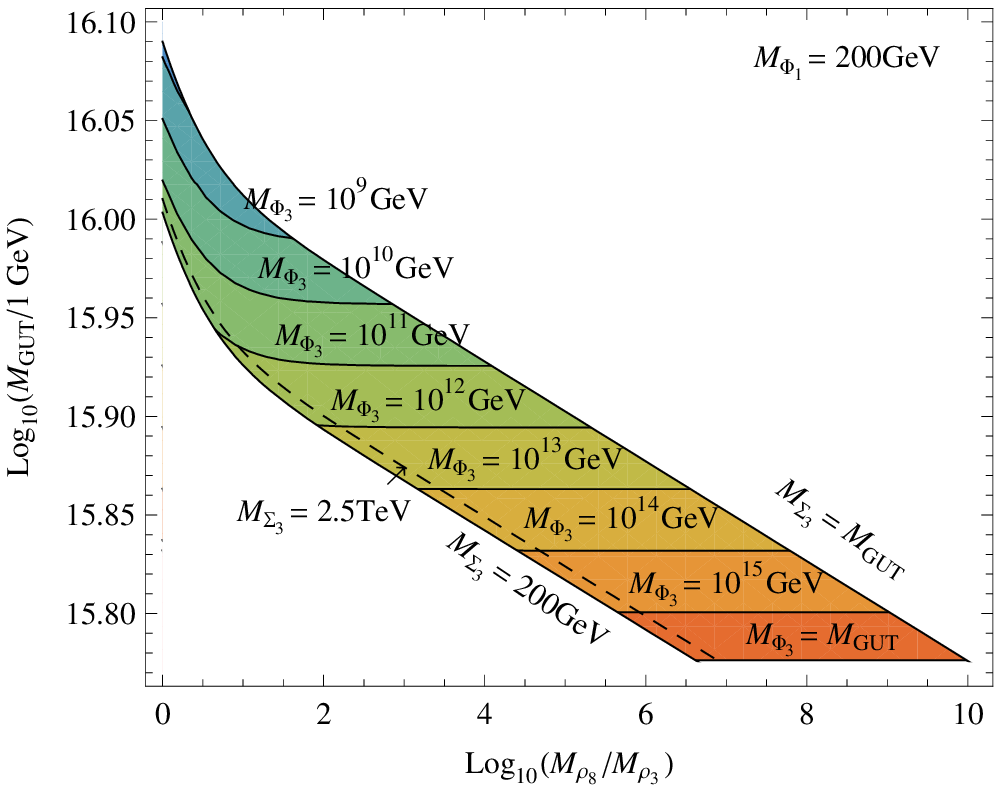}
\caption{Parameter space allowed by unification at one-loop level in
Adjoint $SU(5)$ for the Benchmark I scenario when $M_{\Phi_1}=200$~GeV.}
\label{fig:alpha0Z}
\end{center}
\end{figure}

\begin{figure}[ht]
\begin{center}
\includegraphics[width=9.0cm]{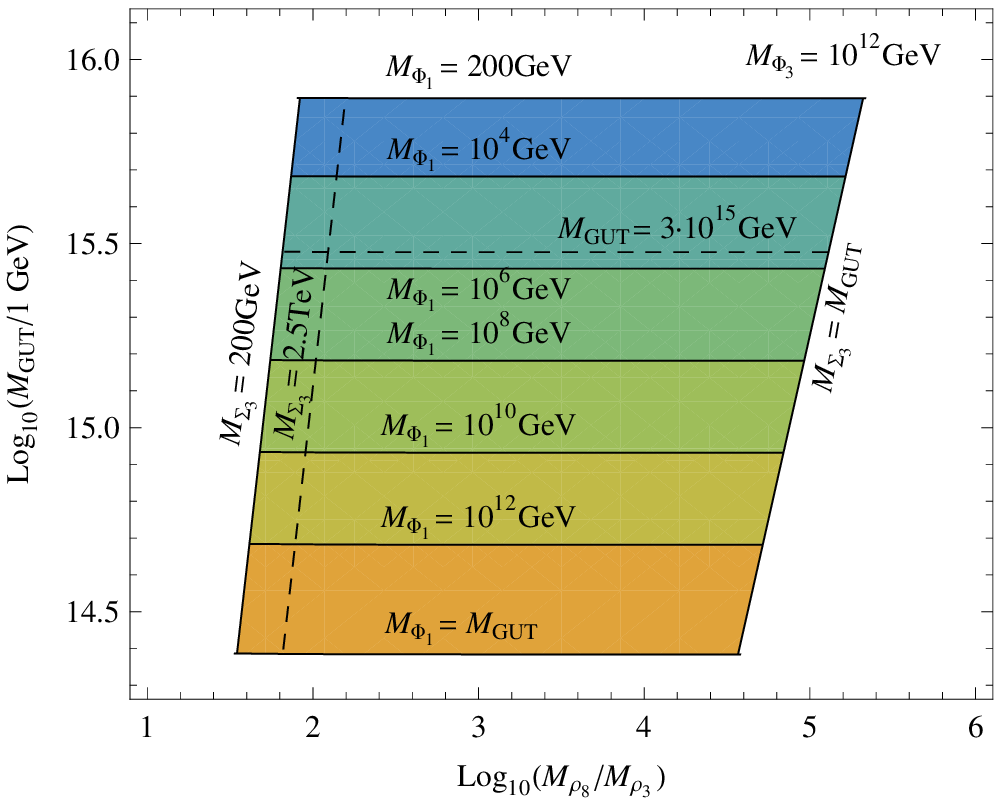}
\caption{Parameter space allowed by unification at one-loop level
in Adjoint $SU(5)$ for the Benchmark I scenario when $M_{\Phi_3}=10^{12}$~GeV.}
\label{fig:alpha3}
\end{center}
\end{figure}

{\it Benchmark I}:
Consider the general case with $\alpha=0$ or $\alpha=2\pi$.
If $m > 3 \, \Lambda$ the lightest field in $\bm{24}$ is $\rho_3$
and $\rho_8$ is the heaviest.
As explained before, the unification will be determined only by the
splitting between the $\rho_8$ and $\rho_3$ masses, and not by their
absolute value, provided that $M_{\rho_8}$ is not larger than $M_{GUT}$
and $M_{\rho_3}$ is above $M_Z$.
The masses of the fields in the $\bm{24}$ multiplet
can be written in terms of one single parameter
$\hat m = M_{\rho_8}/M_{\rho_3}$:
\beq
M_{\rho_0} = \frac15 (3+2 \hat m) \, M_{\rho_3}~, \qquad
M_{\rho_8} = \hat m \, M_{\rho_3}~, \qquad
M_{\rho_{(3,2)}} = M_{\rho_{(\bar{3},2)}} =
\frac12 (1+\hat m) \, M_{\rho_3}~.
\label{mmasses}
\eeq
For $\hat m=1$, or equivalently $\lambda=0$, all masses in the $\rho$
multiplet are equal. For $\hat m \gg 1$ the masses of $\rho_0$, $\rho_8$
and $\rho_{(3,2)}$ are of the same order, and there is a gap between
them and $M_{\rho_3}$.
The solution of the RGEs (\Eq{RGE}) is given by
\bea
M_{GUT} &=& M_Z
\left( \frac{M_{\Sigma_3} \ M_Z^{19}}{M_{\Phi_1}^{20}}\right)^{1/159}
\ \hat{m}^{-12/53}
\
\left( \frac{( 1 + \hat{m})}{2}\right)^{32/159}
\
\exp \left[ \frac{6\pi}{159}
\left( 5 \alpha_1^{-1}+ \alpha_2^{-1} -6  \alpha_3^{-1}\right)(M_Z)\right]
\nn
\\
&=& \frac{M_Z^7}{M_{\Sigma_3} M_{\Phi_3}^5}
\
\left( \frac{\hat{m}^2(1+\hat{m})}{2}\right)^{4/3}
\
\exp \left[ \frac{2\pi}{3}
\left( 5 \alpha_1^{-1}-9 \alpha_2^{-1}
+4 \alpha_3^{-1}\right)(M_Z)\right]
\nn
\\
&=& M_Z \left( \frac{M_Z^5}
{M_{\Phi_1}^4 \ M_{\Phi_3}}\right)^{1/32} \
\left( \frac{1+\hat{m}}{2\hat{m}}\right)^{5/24} \
\exp \left[ \frac{5\pi}{24}
\left( \alpha_1^{-1}-\alpha_3^{-1}\right)(M_Z)\right]~.
\label{RGEsol}
\eea
The available parameter space is shown in Fig.~\ref{fig:alpha0GUT}
and Fig.~\ref{fig:alpha0Z}. In Fig.~\ref{fig:alpha0GUT} we have set $\Phi_1$
at the unification scale, while $M_{\Phi_1}=200$~GeV in Fig.~\ref{fig:alpha0Z}.
We can appreciate that it is possible to achieve unification if both
$\Phi_3$ and $\Sigma_3$ are set at the unification scale when
the splitting of masses is about 7 to 10 orders of magnitude
depending on $M_{\Phi_1}$. The possible solutions shown in
Fig.~\ref{fig:alpha0GUT} are, however, not consistent with
proton decay ($M_{GUT} > 3 \times 10^{15}$ GeV) since the
unification scale is too small. This result constrains
$\Phi_1$ to be below the GUT scale. Notice that
for $M_{\Phi_1}=200$~GeV (Fig.~\ref{fig:alpha0Z}) the unification
scale is larger and consistent with proton decay.
Proton decay constrains also the mass
of $\Phi_3$: $M_{\Phi_3} > 10^{12}$ GeV.
In Fig.~\ref{fig:alpha3} we have set $M_{\Phi_3}$
to $10^{12}$~GeV in order to show the dependence of the unification
scale in terms of $M_{\Phi_1}$. The maximal grand unified scale is
then $M_{GUT}^{max}=7.8 \times 10^{15}$~GeV. For larger values of
$M_{\Phi_3}$ the maximal unification scale will be smaller, and
the parameter space will be more constrained.
Fig.~\ref{fig:alpha3} also tells us that
in order to be consistent with $M_{GUT} > 3 \times 10^{15}$~GeV
the mass of $\Phi_1$ has to be $M_{\Phi_1} < 4.4 \times10^5$~GeV.

\begin{figure}[ht]
\begin{center}
\includegraphics[width=9cm]{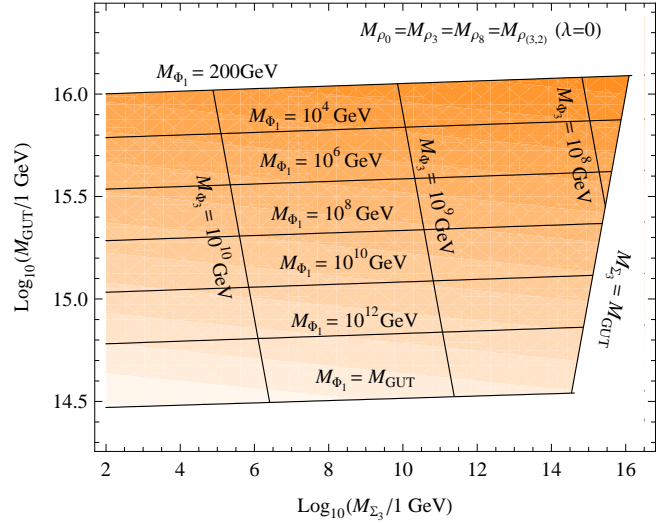}
\caption{\label{fig:lambda0} Parameter space allowed by unification
at one-loop level in Adjoint $SU(5)$ when $\lambda=0$.}
\end{center}
\end{figure}

For completeness, we also show in Fig.~\ref{fig:lambda0} the parameter
space for $\hat m=1$ ($\lambda=0$). This scenario is equivalent to the
case of renormalizable Non-SUSY SU(5) without the contributions of
the fermions in the $\bm{24}$ representation~\cite{Ilja-Pavel-45}.
All the parameter space is however excluded because it requires
$\Phi_3$ to be too light to be consistent with the constraints
discussed before.

\begin{figure}[ht]
\begin{center}
\includegraphics[width=9cm]{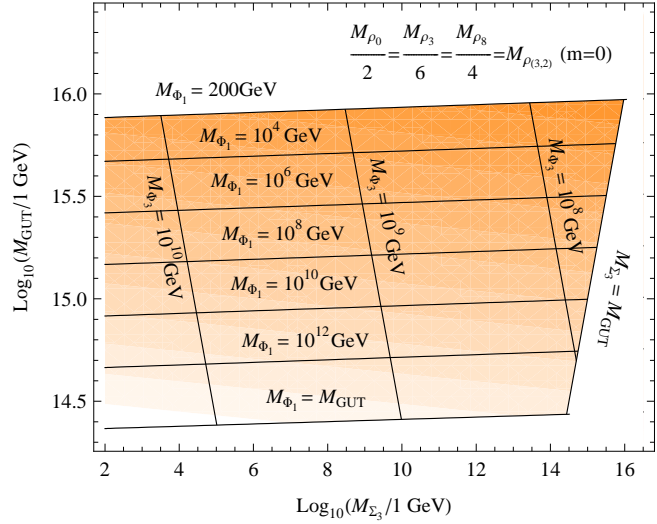}
\caption{\label{fig:m0} Parameter space allowed by unification at
one-loop level in Adjoint $SU(5)$ when $m=0$.}
\end{center}
\end{figure}

{\it Benchmark II}: When $m=0$ the lightest fields in $\bm{24}$
are $\rho_{(3,2)}$ and  $\rho_{(\bar{3},2)}$, and the following
relationship is fulfilled:
\beq
\frac{M_{\rho_0}}{2} = \frac{M_{\rho_3}}{6} =
\frac{M_{\rho_8}}{4} = M_{\rho_{(3,2)}}~.
\eeq
This scenario is illustrated in Fig.~\ref{fig:m0}.
The $\rho$ masses in this scenario are all different
but of the same order of magnitude.
The unification parameter space is therefore quite similar
to the parameter space of Fig.~\ref{fig:lambda0}.
Since $M_{\Phi_3}$ is always below $10^{12}$ GeV, this scenario is ruled
out by proton decay.

{\it Benchmark III}: In the scenario $\alpha= \pi/2$ or $\alpha= 3\pi/2$,
the lightest fields in $\bm{24}$ are $\rho_{(3,2)}$ and $\rho_{(\bar{3},2)}$,
while $\rho_3$ is the heaviest one. The masses are given by:
\bea
M_{\rho_0} &=& \sqrt{ m^2 \ + \ \Lambda^2 }~,
\qquad
M_{\rho_3} = \sqrt{  m^2 \ + \ 9 \ \Lambda^2 }~,
\nn
\\
M_{\rho_8} &=& \sqrt{ m^2 \ + \ 4 \ \Lambda^2 }~,
\qquad
M_{\rho_{(3,2)}} = M_{\rho_{(\bar{3},2)}} = \sqrt{ m^2 \ +
\ \frac{\Lambda^2}{4} }~,
\label{massesalphapi2}
\eea
and can be written in terms of the ratio
$r_{32}=M_{\rho_3}/M_{\rho_{(3,2)}}$:
\beq
M_{\rho_0} = \sqrt{\frac{32+3 \ r_{32}^2}{35}} \, M_{\rho_{(3,2)}}~, \qquad
M_{\rho_3} = r_{32} \, M_{\rho_{(3,2)}}~, \qquad
M_{\rho_8} = \sqrt{\frac{4+3 \ r_{32}^2}{7}} \, M_{\rho_{(3,2)}}~.
\label{mmasses32}
\eeq
For $r_{32}=1$ this scenario is equivalent to the case
$\lambda=0$, and for $r_{32}=6$ to $m=0$.
For a given value of $r_{32}$ in the range $[1,6]$
the parameter space compatible with unification at one-loop
level will interpolate between Fig.~\ref{fig:lambda0} and
Fig.~\ref{fig:m0}. Therefore, this scenario is ruled out by
proton decay.

\begin{figure}[ht]
\begin{center}
\includegraphics[width=8.cm]{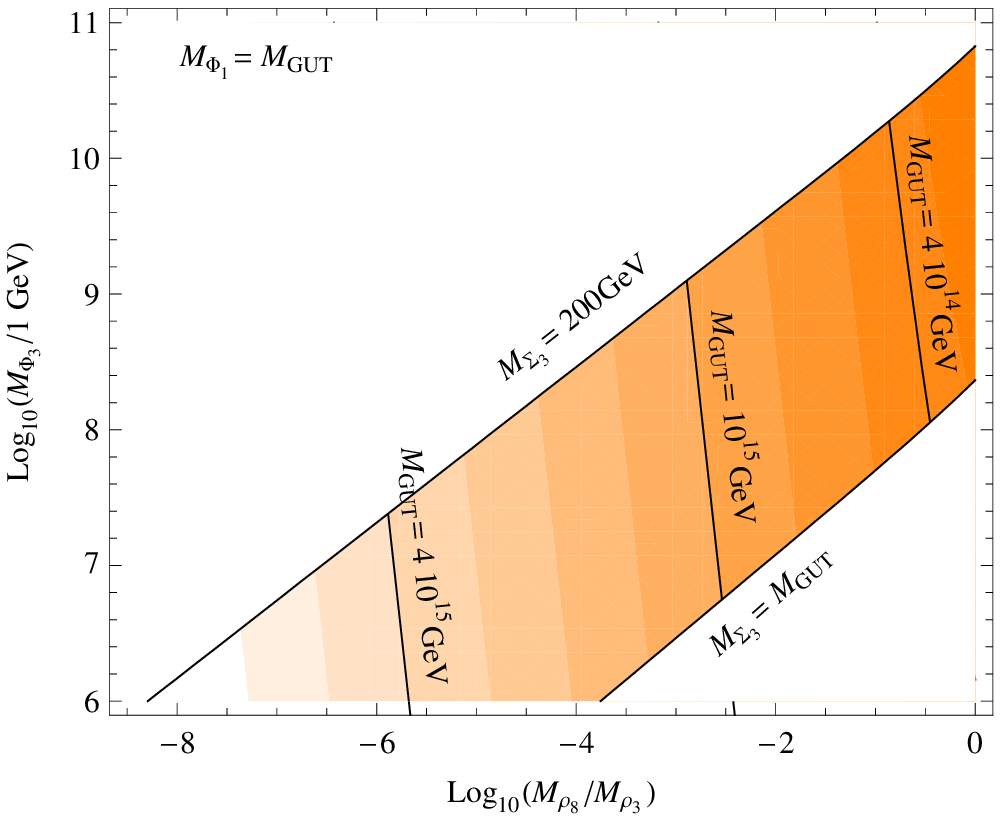}
\includegraphics[width=8.cm]{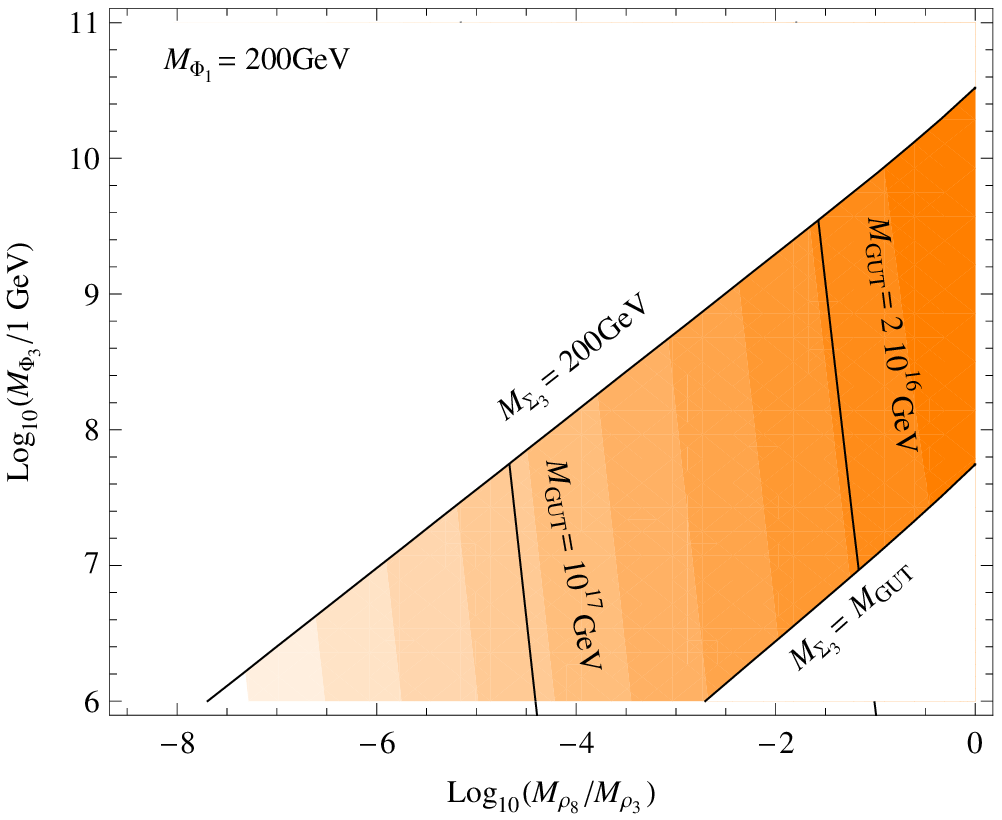}
\caption{\label{fig:alphapi} Parameter space allowed by
unification at one-loop level in Adjoint $SU(5)$ in the Benchmark scenario IV.}
\end{center}
\end{figure}

{\it Benchmark IV}: If $\alpha= \pi$ the heaviest field
in $\bm{24}$ is $\rho_3$ and $\rho_8$ is the lightest
(for $m > 2 \, \Lambda$).
\Eq{mmasses} and \Eq{RGEsol} are valid provided that
$\hat{m}=M_{\rho_8}/M_{\rho_3}$ is smaller
than one. In Fig.~\ref{fig:alphapi} we show the allowed parameter
space for two values of $M_{\Phi_1}$. The parameter space is
however rather insensitive to $M_{\Phi_1}$. If we impose
proton decay this scenario is also excluded.

Let us comment here about the benchmark I scenario with
$0 < m < 3 \, \Lambda$. In this region of the parameter
space $\rho_3$ is not necessary the lightest field, but
all the $\rho$ masses are of the same order, excluding
the points $m \simeq \Lambda$ and $m \simeq \Lambda/2$ where
either $\rho_0$ or $\rho_{(3,2)}$ might be light, respectively.
These scenarios are, however, excluded because they
would require $\Phi_3$ to be too light.
Similar arguments can be employed to exclude the
benchmark IV scenario with $0 < m < 2 \, \Lambda$.

In summary, we have studied the parameter space allowed by unification in
different scenarios. Once the most conservative bounds from the proton
stability are imposed the only scenario which is consistent
with the experiments is the Benchmark scenario I where the
maximal GUT scale is $M_{GUT}^{max} = 7.8 \times 10^{15}$~GeV
for $M_{\Phi_1}=200$~GeV,
and the upper bound on the mass of
the scalar color octet is $M_{\Phi_1} < 4.4 \times 10^{5}$ GeV.
These results are crucial
in order to understand the possibility to test the theory
through proton decay at future experiments.

\section{Cold Dark Matter Constraints}

In supersymmetric theories the lightest supersymmetric
particle, for example the neutralino, is a natural candidate
for the CDM in the Universe once the so-called
R-parity is imposed as a symmetry of the theory.
In non-supersymmetric theories the neutral component
of $\Sigma_3 \sim (\bm{1},\bm{3},0)$ is also a possible candidate
for the CDM, and as we have shown in the previous section
it can be light in Adjoint $SU(5)$. In order to understand this
issue one can imagine a minimal extension of the SM, where
the SM Higgs sector is extended by adding a real triplet $\Sigma_3$.
The matrix representation of $\Sigma_3$ is given by:
\begin{equation}
\Sigma_3 = \frac{1}{2} \left( \begin{array} {cc}
 \Sigma^0  &  \sqrt{2} \Sigma^+ \\
 \sqrt{2} \Sigma^-  & - \Sigma^0
\end{array} \right)~.
\end{equation}
The neutral component of $\Sigma_3$ can be a CDM
candidate if the coupling $H^\dagger \Sigma_3 H$ is not
present and its vacuum expectation value is zero.
This can be achieved by imposing the symmetry
$\Sigma_3 \to - \Sigma_3$. This CDM candidate has been
studied in~\cite{CDM}. As it has been explained by the
authors in~\cite{CDM} the charged $\Sigma^{\pm}$ and
$\Sigma^0$ components have the same mass at tree level.
Once radiative corrections are included, however, 
a mass splitting $\Delta m_{\Sigma}= 166$ MeV is generated, and
the charged component decays mainly through
$\Sigma^+ \to \Sigma^0 \pi^+$.

In our renormalizable GUT theory the 
interaction $H^\dagger_1 \Sigma_3 H_1$ can be eliminated
at tree level by imposing the following condition:
\begin{equation}
\beta_6 = \frac{3}{5} \beta_8 \sqrt{\frac{2}{\pi \alpha_{GUT}}} M_{GUT}~,
\label{fine}
\end{equation}
where $\beta_6$ and $\beta_8$ are, respectively, the couplings
of the $5_H^\dagger 24_H 5_H$ and $5_H^\dagger 24_H^2 5_H$ interactions.
A similar fine-tuning can be made in order to set the couplings
$H_2^\dagger \Sigma_3 H_2$ and $H_2^\dagger \Sigma_3 H_1$ to zero.
See Appendix A for all relevant scalar interactions. 
In this GUT theory, however, the symmetry $\Sigma_3 \to - \Sigma_3$ 
cannot be realized without embedding in  ${\bf 24}_H \to - {\bf 24}_H$ at the GUT level,
which prevents to achieve unification in agreement with the experiment.
In this case all the fields in the fermionic {\bf 24} representation 
would have the same mass and would not contribute to the running 
of the gauge couplings at one-loop. Also the field $\Sigma_3$ would not 
be neither light in this case.
The fine-tuning in \Eq{fine} can thus not be made stable under radiative 
corrections, and the suppression of the undesired interactions 
is possible only order by order. 
Although, this fact makes less appealing the idea of $\Sigma^0$ as 
a CDM candidate, in our opinion, it deserves some attention
because the fine tuning is always possible.  

\begin{figure}[ht]
\begin{center}
\includegraphics[width=9.0cm]{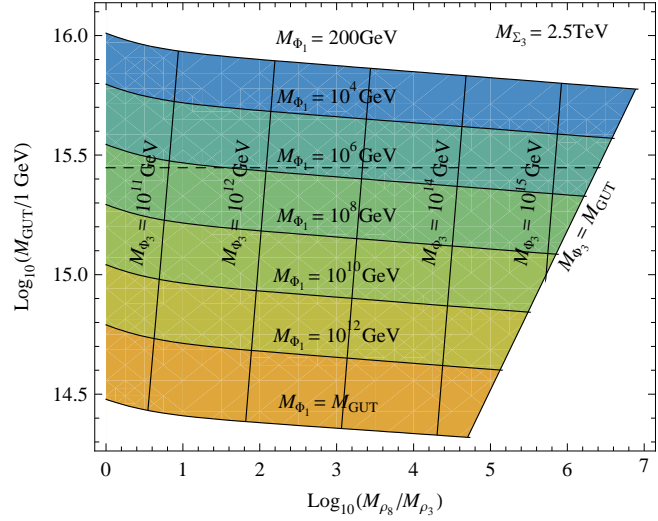}
\caption{Parameter space allowed by unification at one-loop level
in Adjoint $SU(5)$ for the Benchmark I scenario when
$M_{\Sigma_3}=2.5$~TeV. Dashed line at $M_{GUT}=3\times 10^{15}$~GeV.}
\label{fig:alpha4}
\end{center}
\end{figure}

It has been pointed out in~\cite{CDM} that if the mass of the neutral component
of the real triplet is $M_{\Sigma_3}\approx 2.5$ TeV the thermal relic abundance
is equal to the observed Dark Matter (DM) abundance $\Omega_{DM} h^2 = 0.110 \pm 0.005$.
Therefore, if we stick to the possibility of having $\Sigma^0$ as a CDM candidate
one could say that only the region of the parameter space consistent with unification
shown in the previous section where $M_{\Sigma_3}\approx 2.5 $ TeV is allowed by
the Dark Matter constraints. We show this case in Fig.~\ref{fig:alpha4}.

\section{Upper Bounds on the Partial Proton Decay Lifetimes}
As we have mentioned in the previous sections there
are five fields in this theory that mediate proton decay.
These are the superheavy gauge
bosons $V \sim (\bm{3},\bm{2},-5/6)\bigoplus(\overline{\bm{3}},\bm{2},5/6)$,
and the $SU(3)$ triplets $T$, $\Phi_3$, $\Phi_5$ and $\Phi_6$. The least model
dependent and usually the dominant proton decay contribution in
non-supersymmetric scenarios comes from gauge boson mediation.
It is important to understand the possibility to test this theory
at future proton decay experiments~\cite{Experiments}.
Assuming that the Yukawa matrix for up quarks is symmetric,
$Y_U=Y_U^T$, the golden channels to test the theory are~\cite{test1}:
\begin{eqnarray}
\Gamma(p \rightarrow \pi^+ \bar{\nu}) &=&
\sum_{i=1}^3 \ \Gamma(p \to \pi^+\bar{\nu}_i)=
\frac{\pi}{2} \frac{\alpha_{GUT}^2}{M_{GUT}^4} \frac{m_p}{f_{\pi}^2} \
|V_{ud}|^2 \ A_r^2 \ |\alpha|^2
\left( 1 \ + \ D \ + \ F \right)^2~,
\end{eqnarray}
and
\begin{eqnarray}
\Gamma(p \to K^+\bar{\nu})&=&
\sum_{i=1}^3 \ \Gamma(p \to K^+\bar{\nu}_i) =
\frac{\pi}{2} \frac{\alpha_{GUT}^2}{M_{GUT}^4}
\frac{(m_p^2 - m_K^2)^2}{m_p^3 \ f_{\pi}^2} \ A_r^2 \ |\alpha|^2
\left( A_1^2 \ |V_{ud}|^2 \ + \ A_2^2 \ |V_{us}|^2 \right)~, \nn \\
\end{eqnarray}
where
\begin{eqnarray}
A_1&=& \frac{2 m_p }{3 m_B} D, \qquad \text{and} \qquad A_2
= 1 + \frac{m_p}{3 m_B}(D + 3F)~.
\end{eqnarray}
In the above expressions $D$ and $F$ are the parameters of the
Chiral Lagrangian~\cite{WiseCL,review} and $m_B =1.150 \ \text{GeV} \approx m_{\Sigma} \approx m_{\Lambda}$
is the averaged Baryon mass. The values $F=0.463$ and $D=0.804$
~\cite{Cabibbo} are used as input values for our numerical analysis,
$\alpha$ is the matrix element of the three quark states between
the proton and the vacuum state: $\langle 0| \epsilon_{abc} \, \epsilon_{\alpha\beta} \, u_{aR}^{\alpha}
\, d_{bR}^{\beta} \, u_{L}^{\gamma} | p \rangle = \alpha \, u_{L}^{\gamma}$
(we use $\alpha=0.015$ GeV$^3$~\cite{lattice} calculated at the scale 2.3 GeV), and $A_r=A_L A_S$ is the
renormalization factor. See Ref.~\cite{review} for more details.
The long-range renormalization factor is given by
\begin{equation}
A_L = \left( \frac{\alpha_3 (m_b)}{ \alpha_3 (M_Z)} \right)^{6/23}
\left( \frac{\alpha_3 (Q)}{ \alpha_3 (m_b)} \right)^{6/25}
\approx 1.25 \qquad \text{using} \qquad Q=2.3 \ \text{GeV},
\end{equation}
while the short distance renormalization factor reads as
\begin{equation}
A_S = \left( \frac{\alpha_3 (M_{GUT})}{ \alpha_3 (M_Z)} \right)^{2/b_3}.
\end{equation}
Using the maximal allowed value for the GUT scale
$M_{GUT}^{max}=7.8 \times 10^{15}$ GeV,
which corresponds to set $M_{\Phi_3}=10^{12}$~GeV and $M_{\Phi_1}=200$~GeV
(This value is almost independent of $\hat{m}$ and $M_{\Sigma_3}$.),
we find the following upper bounds on the proton decay
partial lifetimes
\begin{eqnarray}
\tau \left( p \to K^+ \bar{\nu} \right) & \leq & 9.3 \times 10^{36} \ \text{years} \qquad \text{and} \qquad
\tau \left( p \to \pi^+ \bar{\nu} \right) \leq 3.0 \times 10^{35} \ \text{years},
\end{eqnarray}
where $\alpha_{GUT}^{-1}=33.4$, which corresponds to
set $M_{\rho_8}=M_{GUT}$. The present experimental lower bounds
are $\tau^{exp} \left( p \to K^+ \bar{\nu} \right) >
2.3 \times 10^{33}$ years~\cite{lowerbounds}
and  $\tau^{exp} \left( p \to \pi^+ \bar{\nu} \right) >
2.5 \times 10^{31}$ years~\cite{PDG},
respectively. Future experiments~\cite{Experiments}
will improve these bounds by two or more orders of magnitude.
We expect therefore that this theory could be tested
at the next proton decay experiments.
It is important to say that if one neglects
the mixing between quarks and leptons one gets similar
predictions for all proton decay channels.
Particularly, one gets $\tau (p \to e^+ \pi^0)\leq 1.2 \times 10^{35}$
years.

\section{Light Exotic Fields: Colored Scalar Octets}

The phenomenological aspects of an extension of the SM where the 
Higgs sector is composed of the SM Higgs and a scalar color octet have been
studied recently in Ref.~\cite{Wise}. The authors in~\cite{Wise}
have noticed that the color octet has Yukawa couplings to quarks
with natural flavour conservation. After the electroweak symmetry breaking
there are four physical Higgses in the adjoint of $SU(3)$, $S_R^0$
(Real component of $S^0$), $S_I^0$ (Imaginary component of $S^0$),
and $S^{\pm}$. As expected, the splitting between their masses
is of order $M_W$. The different decays of the octets, the pair and
singlet production mechanisms at the LHC have been studied in
detail in~\cite{Wise}. For other studies see also~\cite{Pheno-Octets}.
The presence of these light exotic fields will also modify the Higgs
production at the LHC and its decays~\cite{Octet-Higgs}.

The Adjoint $SU(5)$ GUT theory predicts the existence of these light 
exotic fields in a natural way; this is the field $\Phi_1 \sim (\bm{8},\bm{2},1/2)$.
As we have discussed in the previous sections by imposing all relevant 
constraints from the unification of gauge interactions and proton decay 
we have found that the upper bound on the mass of this color octet 
is $M_{\Phi_1} < 4.4 \times 10^5$~GeV. 
The lightness of $\Phi_1$ is needed for achieving a GUT scale 
large enough to be in agreement with proton decay. 
Following the notation of~\cite{Wise}:
\begin{equation}
\Phi_1 =
\left(
\begin{array} {c}
 S^+  \\
 S^0
\end{array} \right) = S^a T^a,
\end{equation}
where $a=1,..,8$ and $T^a$ are the $SU(3)$ generators.
This leads to very exciting phenomenological implications
and the possibility to test the theory at the LHC.
The properties of the light colored octets in this context will
be studied in detail in a future publication.

\section{Summary and Outlook}
We have investigated the unification of gauge
interactions in the context of Adjoint SU(5). In this simple GUT
theory the neutrino masses are generated through the Type I and Type III
seesaw mechanisms. We have found the following phenomenological 
predictions:

\begin{itemize}

\item Among all the possible mass spectra of the theory, 
there is only one scenario for the relation between
the masses of the fermionic fields in the adjoint representation which
is consistent with unification of gauge couplings and the conservative
bounds coming from proton decay. In the allowed parameter space the lightest
field is $\rho_3$. This result is crucial to understand the Baryogenesis
via Leptogenesis mechanism in this context.

\item In this theory, we have identified the neutral component of the real scalar
triplet, $\Sigma_3 \sim (\bm{1},\bm{3},0)$, as a possible CDM candidate.
This possibility requires, however, a delicate fine-tuning to suppress 
the interaction of $\Sigma_3$ to other color singlet Higgses,
which cannot be protected by a discrete symmetry because
$\Sigma_3$, in the GUT theory, is embedded in a larger multiplet. 

\item Since the predicted upper bounds on the proton decay partial lifetimes
are $\tau( p \to K^+ \bar{\nu}) \ \leq \ 9.3 \times 10^{36}\ $ years and
$\tau( p \to \pi^+ \bar{\nu}) \ \leq \ 3.0 \times 10^{35}$ years,
future proton decay experiments could test this theory.

\item The theory predicts light colored scalar octets,
$S_R^0$, $S_I^0$, and $S^{\pm}$, which could be produced at the 
Large Hadron Collider and modify the production and decays of the SM Higgs.
The properties of these fields in this context will be studied in a 
future publication.

\end{itemize}

\begin{acknowledgments}
We would like to thank Manuel Drees for the careful reading of the
manuscript and very useful comments. The work of P. F. P. was supported
in part by the U.S. Department of Energy contract No. DE-FG02-08ER41531
and in part by the Wisconsin Alumni Research Foundation. P. F. P. would
like to thank S. Blanchet for discussions. The work of G.R. was
supported by Ministerio de Ciencia e Innovaci\'on under
grants FPA2007-60323 and CSD2007-00042, and European Commission
MRTN FLAVIAnet under contract MRTN-CT-2006-035482.
\end{acknowledgments}

\appendix
\section{Field Content and Interactions in Adjoint SU(5)}
In Adjoint $SU(5)$~\cite{adjoint} the matter is unified in
\begin{eqnarray}
{\bm {\overline 5}} &=& l_L \bigoplus (d^C)_L = ({\bm 1},{\bm 2},-1/2)
\bigoplus ({\bm {\overline 3}},\bm{1},1/3),
\\
{\bm {10}} &=& (u^C)_L \bigoplus q_L \bigoplus (e^C)_L
= ({\bm {\overline{3}}},\bm{1},-2/3) \bigoplus ({\bm 3},{\bm 2},1/6)
\bigoplus ({\bm 1},\bm{1},1),
\end{eqnarray}
and
\begin{eqnarray}
{\bm 24} &=& (\rho_8)_L \bigoplus (\rho_3)_L \bigoplus (\rho_{(3,2)})_L
\bigoplus (\rho_{(\bar{3}, 2)})_L \bigoplus (\rho_{0})_L \nonumber \\
&=& ({\bm 8},{\bm 1},0) \bigoplus ({\bm 1},{\bm 3},0) \bigoplus ({\bm 3},{\bm 2},-5/6)
\bigoplus ({\bm {\overline 3}},{\bm 2},5/6) \bigoplus ({\bm 1},{\bm 1},0),
\end{eqnarray}
while the Higgs sector is composed of
\begin{eqnarray}
\bm{5}_H&=& H_1 \bigoplus T = (\bm{1},\bm{2},1/2)
\bigoplus (\bm{3},\bm{1},-1/3),\\
\bm{24}_H&=&\Sigma_8 \bigoplus \Sigma_3 \bigoplus \Sigma_{(3,2)}
 \bigoplus
\Sigma_{(\overline{3},2)} \bigoplus \Sigma_{24} \nonumber \\
&=& (\bm{8},\bm{1},0) \bigoplus (\bm{1},\bm{3},0) \bigoplus
(\bm{3},\bm{2},-5/6) \bigoplus (\overline{\bm{3}},\bm{2},5/6)
\bigoplus (\bm{1},\bm{1},0),
\end{eqnarray}
and
\begin{eqnarray}
\bm{45}_H&=&\Phi_1 \bigoplus \Phi_2 \bigoplus \Phi_3
\bigoplus \Phi_4 \bigoplus \Phi_5 \bigoplus \Phi_6 \bigoplus H_2 \nonumber \\
&=& (\bm{8},\bm{2},1/2) \bigoplus (\overline{\bm{6}},\bm{1}, -1/3)
 \bigoplus
(\bm{3},\bm{3},-1/3) \bigoplus (\overline{\bm{3}}, \bm{2}, -7/6) \nonumber \\
&\bigoplus& (\bm{3},\bm{1}, -1/3) \bigoplus (\overline{\bm{3}}, \bm{1},
 4/3)
\bigoplus (\bm{1}, \bm{2}, 1/2).
\end{eqnarray}
The Action of this model is given by
\begin{eqnarray}
S = S_{\rm kinetic}  + S_{\rm Yukawa} + S_{\rm scalar},
\end{eqnarray}
\begin{eqnarray}
S_{\rm kinetic} &=& \int d^4x~[-\frac{1}{4} {\rm Tr} F^{\mu\nu}F_{\mu\nu}
                + \frac{1}{2} (D^{\mu} 5_H)^{\dagger} (D_{\mu} 5_H) +
                \\ \nonumber
                &+& {\rm Tr}(D^{\mu} 24_H)^{\dagger} (D_{\mu} 24_H)
                +    {{\rm Tr}(D^{\mu} 45_H)^{\dagger}
                 (D_{\mu} 45_H)} +
                \\ \nonumber
                &+& \bar{5}^{\dagger} \gamma^0 i \gamma^{\mu} D_{\mu} \bar{5}
                \ + \ {\rm Tr}~(\overline{10} i\gamma^{\mu} D_{\mu} 10)~
\ + \ {\rm Tr}~(\overline{24} i\gamma^{\mu} D_{\mu} 24)~],
\end{eqnarray}
where
\begin{eqnarray}
D_{\mu} 5_H &=& \partial_{\mu} 5_H + i g_{\small GUT} A_{\mu} 5_H,
             ~~~~
D_{\mu} 10 =  \partial_\mu 10 \ + \ i g_{\small GUT} \left( A_\mu 10 \ + \ 10 A_\mu^T \right),
\\
D_{\mu}\bar{5} &=& \partial_{\mu}\bar{5} \ - \ i g_{\small GUT} \ A^T_{\mu} \bar{5},
              ~~~~~
D_{\mu} 24_H = \partial_{\mu} 24_H \ + \ i g_{\small GUT} [A_{\mu}, 24_H],
\\
D_{\mu} (45_H)^{\alpha \beta}_{\gamma} &=& \partial_\mu (45_H)^{\alpha \beta}_{\gamma}
\ + \ i g_{\small GUT} \left( A_\mu^{\alpha m} (45_H)^{m \beta}_{\gamma} \ + \
(45_H)^{\alpha m}_\gamma (A_\mu^T)^{m \beta} \ - \
(A_\mu^T)_{\gamma \delta} (45_H)^{\alpha \beta}_{\delta}\right),\nonumber
\\
\\
F_{\mu\nu} &=& \partial_{\mu} A_{\nu} - \partial_{\nu} A_{\mu}
           - i g_{\small GUT} [A_{\mu}, A_{\nu}],
\end{eqnarray}
\begin{eqnarray}
S_{\rm Yukawa} &=& -
\int d^4 x
\left( Y_1^{a b} 10^{ij}_a ~\bar{5}^i_b ~(5^*_H)^j
\ + \
Y_2^{a b} 10^{ij}_a ~\bar{5}^l_b ~(45^*_H)^{ij}_l \right)
\nonumber \\
&-& \int d^4 x~ \epsilon_{ijklr}~\left(~ Y_3^{a b} 10^{ij}_a~ 10^{kl}_b~
5^r_H  \
+ \
Y_4^{a b} 10^{ij}_a ~10^{pk}_b ~(45_H)^{lr}_p \right)
\nonumber \\
&-&  \int d^4 x \left( c_a \ \bar{5}_{a i} 24^i_{~k} 5_H^k \ + \
p_a \ \bar{5}_{a i} 24^j_{~k} (45_H)^{ik}_j \right) \ + \ S_{24} \ + \ \text{h.c.}
\label{Yukawas}
\end{eqnarray}
The additional terms relevant for the seesaw mechanism are given by
\begin{eqnarray}
S_{24} &=&  - \int d^4 x \left( M \ 24^i_{~j} 24^j_{~i}
\ + \ \lambda \ 24^i_{j~} 24^j_{~k} (24_H)^k_{~i} \right) \ + \ \text{h.c.}
\label{seesaw2}
\end{eqnarray}
In order to simplify our notation in the next equations
we use $5_H=5$, $24_H=24$ and $45_H=45$. The scalar interactions
are given by
\begin{eqnarray}
S_{\rm scalar} = - \int d^4 x \ V(5_H, 24_H, 45_H).
\end{eqnarray}
The field $45_H$ satisfies the following conditions:
$(45_H)^{ij}_k = -(45_H)^{ji}_k$, $\Sigma^5_{\i=1}(45_H)^{ij}_i =0$,
and
$v_{45}/\sqrt{2} = \langle 45_H\rangle^{15}_1 = \langle 45_H\rangle^{25}_2
       = \langle 45_H\rangle^{35}_3$. The $SU(5)$ invariant
Higgs potential for our model is
\begin{eqnarray}
V(24_H, 45_H, 5_H)& = & V_1(24_H) + V_2(45_H) + V_3(5_H)
                  + V_4(24_H, 45_H) \\ \nonumber &+& V_5(24_H, 5_H)
                   + V_6(5_H, 45_H) + V_7(24_H, 45_H, 5_H),
\end{eqnarray}
where
\begin{eqnarray}
V_1(24_H) &=&-\frac{\mu^2_{24}}{2} 24^\alpha_{~\beta} 24^\beta_{~\alpha}
\ + \
\frac{a_{1}}{2}
(24^\alpha_{~\beta} 24^\beta_{~\alpha})^2
\ + \
\frac{a_{2}}{3} 24^\alpha_{~\beta} 24^\beta_{~\gamma} 24^\gamma_{~\alpha}
\ + \
\frac{a_{3}}{2} 24^\alpha_{~\beta} 24^\beta_{~\gamma} 24^\gamma_{~\delta} 24^\delta_{~\alpha},
\end{eqnarray}
\begin{eqnarray}
V_2(45_H)
&=&
-\frac{1}{2} \mu^2_{45} \ ({45}^{\alpha \beta}_\gamma {45}^\gamma_{\alpha \beta})
\ + \
\lambda_1 \ (45^{\alpha \beta}_\gamma 45^\gamma_{\alpha \beta})^2
\ + \
\lambda_2 \
45^{\alpha \beta}_\gamma 45^\delta_{\alpha \beta} 45^{k \lambda}_\delta 45^\gamma_{k \lambda} +
\nonumber \\
&+& \
\lambda_3 \ 45^{\alpha \beta}_\gamma 45^\delta_{\alpha \beta}
45^{k \gamma}_{\lambda} 45^{\lambda}_{k \delta}
\ + \
\lambda_4 \ 45^{\alpha \delta}_\beta 45^\beta_{\alpha \gamma} 45^{k \gamma}_{\lambda} 45^{\lambda}_{k \delta}
\ + \
\lambda_5 \ 45^{\alpha \gamma}_\delta 45^\beta_{\gamma \epsilon} 45^{k \delta}_\alpha
45^\epsilon_{k\beta} +
\nonumber \\
& + &
\lambda_6 \ 45^{\alpha \gamma}_\delta 45^\beta_{\gamma \epsilon} 45^{k \epsilon}_\alpha 45^\delta_{k \beta}
\ + \
\lambda_7 \ 45^{\alpha \gamma}_\delta 45^\beta_{\gamma \epsilon} 45^{k \delta}_\beta 45^\epsilon_{k \alpha}
\ + \
\lambda_8 \ 45^{\alpha \gamma}_\delta 45^\beta_{\gamma \epsilon} 45^{k \epsilon}_\beta 45^\delta_{k \alpha}.
\nonumber \\
\end{eqnarray}
See reference~\cite{Frampton}. The rest of the scalar interactions are given by
\begin{eqnarray}
V_3(5_H) = -\frac{\mu^2_5}{2} \ 5^*_\alpha 5^\alpha \
+ \ \frac{a_4}{4} \ \left( 5^*_\alpha 5^\alpha \right)^2,
\end{eqnarray}
\begin{eqnarray}
V_4({24_H, 45_H})
&=&
a_5 \ 45^{\alpha \beta}_\gamma 24^\gamma_{~\delta} 45^\delta_{\alpha \beta}
\ + \ a_6 \
(45^{\alpha \beta}_\gamma 45^\gamma_{\alpha \beta}) 24^\delta_{~\epsilon} 24^\epsilon_{~\delta}
\ + \ \beta_1 \
45^{\alpha \beta}_\gamma 24^\delta_{~\alpha} 24^\epsilon_{~\beta} 45^\gamma_{\delta \epsilon} +
\nonumber \\
& + &  \beta_2
\ 45^{\alpha \beta}_\gamma 24^\gamma_{~\beta} 24^\delta_{~\epsilon} 45^\epsilon_{\alpha \delta}
\ + \ \beta_3
\ 45^{\alpha \beta}_\gamma 24^\gamma_{~\epsilon} 24^\delta_{~\beta} 45^\epsilon_{\alpha \delta}
\ + \ \beta_4
\ 45^{\alpha \beta}_\gamma 24^k_{~\alpha} 24^\lambda_{~k} 45^\gamma_{\lambda \beta} +
\nonumber \\
& + & \ \beta_5
\ 45^{\alpha \beta}_\gamma 24^\gamma_{~k} 24^k_{~\lambda} 45^\lambda_{\alpha \beta},
\end{eqnarray}
\begin{eqnarray}
V_5(24_H, 5_H) =
\beta_6 \ 5^*_\alpha 24^\alpha_{~\beta} 5^\beta
\ + \
\beta_7 \ 5^*_\alpha 5^\alpha 24^\beta_{~\gamma} 24^\gamma_{~\beta}
\ + \
\beta_8 \ 5^*_\alpha 24^\alpha_{~\beta} 24^\beta_{~\gamma} 5^\gamma,
\end{eqnarray}
\begin{eqnarray}
V_6(5_H, 45_H) =
c_1 \ (45^{\alpha \beta}_\gamma 45^\gamma_{\alpha \beta}) 5^*_\delta 5^\delta
\ + \
c_2 \ 45^{\alpha \beta}_\delta \ 5^*_\gamma \ 45^\gamma_{\alpha \beta} \ 5^\delta
\ + \
c_3 \ 45^{\alpha \beta}_{\gamma} \ 45^{\gamma}_{\alpha \delta} \ 5^*_\beta \ 5^\delta,
\end{eqnarray}
and
\begin{eqnarray}
V_7(24_H, 45_H, 5_H) = c_4 \ 5^*_{\alpha} 24^{\gamma}_{~\beta} 45^{\alpha \beta}_{\gamma} \ +
\ c_5 \ 5^*_{\alpha} 24^{\gamma}_{~\delta} 24^{\delta}_{~\beta} 45^{\alpha \beta}_{\gamma} \ + \
\ c_6 \ 5^*_{\alpha} 24^{\alpha}_{~\beta} 24^{\gamma}_{~\delta} 45^{\beta \delta}_{\gamma} \ + \
\text{h.c.}
\end{eqnarray}
Notice that we have generalized the results for the scalar potential
presented in references~\cite{Pkalyniak} and \cite{Peckert}.

\section{Running of gauge couplings}
In order to understand the predictions coming
from the unification of gauge couplings at the
scale $\Lambda=M_{GUT}$ one uses the RGEs:
\begin{eqnarray}
\frac{1}{\alpha_i (M_Z)} - \frac{1}{\alpha_i (\Lambda)} & = &
\frac{1}{2\pi} \ b_i^{SM} \ln \frac{\Lambda}{M_Z} \ +
\ \frac{1}{2\pi} \sum_I \ b_{iI} \ \Theta \left( \Lambda - M_I \right) \ln \frac{\Lambda}{M_I},
\end{eqnarray}
where the function $\Theta (x)$ is one for $x>0$ and zero for $x \leq 0$. The different
contributions to the running of the gauge couplings are listed in Table~\ref{contributions}.
\begin{table}[ht]
\caption{\label{contributions} Contributions to the running of gauge couplings.}
\begin{ruledtabular}
\begin{tabular}{lcccc}
 \text{Fields} & $b_1$ & $b_2$ & $b_3$
\\
\hline
$H_1$ &  1/10 & 1/6 & 0
\\
\hline
$T$ &  1/15 & 0 & 1/6
\\
\hline
$\Sigma_8$ &  0 & 0 & 1/2
\\
\hline
$\Sigma_3$ &  0 & 1/3 & 0
\\
\hline
$\Phi_{1}$ &  4/5 & 4/3 & 2
\\
\hline
$\Phi_{2}$ &  2/15 & 0 & 5/6
\\
\hline
$\Phi_{3}$ &  1/5 & 2 & 1/2
\\
\hline
$\Phi_{4}$ & 49/30 &  1/2 & 1/3
\\
\hline
$\Phi_{5}$ & 1/15 & 0 & 1/6
\\
\hline
$\Phi_{6}$ &  16/15 & 0 & 1/6
\\
\hline
$H_2$ & 1/10 & 1/6 & 0
\\
\hline
$\rho_8$ &  0 & 0 & 2
\\
\hline
$\rho_3$ & 0 & 4/3 & 0
\\
\hline
$\rho_{(3,2)}$, $\rho_{(\bar{3},2)}$  & 5/3 & 1 & 2/3
\end{tabular}
\end{ruledtabular}
\end{table}
$b_1^{SM}=41/10$, $b_2^{SM}=-19/6$ and $b_3^{SM}=-7$.



\begin{thebibliography}{99}


\bibitem{GW}
  H.~Georgi, H.~R.~Quinn and S.~Weinberg,
  ``Hierarchy Of Interactions In Unified Gauge Theories,''
  Phys.\ Rev.\ Lett.\  {\bf 33} (1974) 451.

\bibitem{GG}
H.~Georgi and S.~L.~Glashow,
``Unity Of All Elementary Particle Forces,''
Phys.\ Rev.\ Lett.\  {\bf 32} (1974) 438.

\bibitem{PS}
  J.~C.~Pati and A.~Salam,
  ``Is Baryon Number Conserved?,''
  Phys.\ Rev.\ Lett.\  {\bf 31} (1973) 661.

\bibitem{PDG}
W.-M. Yao {\it et al.},
Journal of Physics, {\bf G} 33, 1 (2006).

\bibitem{Experiments}
  D.~Autiero {\it et al.},
  ``Large underground, liquid based detectors for astro-particle physics in
  Europe: scientific case and prospects,''
  JCAP {\bf 0711} (2007) 011
  [arXiv:0705.0116 [hep-ph]];
  T.~M.~Undagoitia {\it et al.},
  ``Search for the proton decay p to K+ anti-nu in the large liquid
  scintillator low energy neutrino astronomy detector LENA,''
  Phys.\ Rev.\  D {\bf 72} (2005) 075014
  [arXiv:hep-ph/0511230];
  A.~Bueno {\it et al.},
  ``Nucleon decay searches with large liquid argon TPC detectors at shallow
  depths: Atmospheric neutrinos and cosmogenic backgrounds,''
  JHEP {\bf 0704} (2007) 041
  [arXiv:hep-ph/0701101];
  M.~Diwan {\it et al.},
  ``Proposal for an experimental program in neutrino physics and proton  decay
  in the homestake laboratory,''
  arXiv:hep-ex/0608023;
  A.~de Bellefon {\it et al.},
  ``MEMPHYS: A large scale water Cerenkov detector at Frejus,''
  arXiv:hep-ex/0607026.

\bibitem{review}
  P.~Nath and P.~Fileviez P\'erez,
  ``Proton stability in grand unified theories, in strings, and in branes,''
  Phys.\ Rept.\  {\bf 441} (2007) 191
  [arXiv:hep-ph/0601023].

\bibitem{GJ}
  H.~Georgi and C.~Jarlskog,
  ``A New Lepton - Quark Mass Relation In A Unified Theory,''
  Phys.\ Lett.\ B {\bf 86} (1979) 297.

\bibitem{TypeI}
  P.~Minkowski,
  ``Mu $\to$ E Gamma At A Rate Of One Out Of 1-Billion Muon Decays?,''
  Phys.\ Lett.\ B {\bf 67} (1977) 421 ;
  T. Yanagida, in {\it Proceedings of the Workshop on the Unified Theory
   and the Baryon Number in the Universe}, eds. O. Sawada et al., (KEK
   Report~79-18, Tsukuba, 1979), p.~95;
  M. Gell-Mann, P. Ramond and R. Slansky,
   in {\it Supergravity}, eds. P. van Nieuwenhuizen et al.,
   (North-Holland, 1979), p.~315;
  S.L. Glashow, in {\it Quarks and Leptons}, Carg\`ese, eds. M. L\'evy et al.,
(Plenum, 1980), p. 707;
  R.~N.~Mohapatra and G.~Senjanovi\'c,
  ``Neutrino Mass And Spontaneous Parity Nonconservation,''
  Phys.\ Rev.\ Lett.\  {\bf 44} (1980) 912.

\bibitem{Ksbabu}
  K.~S.~Babu and E. Ma
  ``Suppression of proton decay in $SU(5)$ grand unification,''
   Phys. \ Lett. \ B {\bf 144}, 381 (1984).

\bibitem{Ilja-Pavel-45}
  I.~Dorsner and P.~Fileviez~P\'erez,
  ``Unification versus proton decay in SU(5),''
  Phys.\ Lett.\  B {\bf 642} (2006) 248
  [arXiv:hep-ph/0606062];
  I.~Dorsner and I.~Mocioiu,
  ``Predictions from type II see-saw mechanism in SU(5),''
  Nucl.\ Phys.\  B {\bf 796} (2008) 123
  [arXiv:0708.3332 [hep-ph]].



\bibitem{TypeII}
W.~Konetschny and W.~Kummer,
``Nonconservation Of Total Lepton Number With Scalar Bosons,''
  Phys.\ Lett.\  B {\bf 70} (1977) 433;
T.~P.~Cheng and L.~F.~Li,
  ``Neutrino Masses, Mixings And Oscillations In SU(2) X U(1) Models Of
  Electroweak Interactions,''
  Phys.\ Rev.\  D {\bf 22} (1980) 2860;
  G.~Lazarides, Q.~Shafi and C.~Wetterich,
 ``Proton Lifetime And Fermion Masses In An SO(10) Model,''
  Nucl.\ Phys.\ B {\bf 181} (1981) 287;
  J.~Schechter and J.~W.~F.~Valle,
  ``Neutrino Masses In SU(2) X U(1) Theories,''
  Phys.\ Rev.\ D {\bf 22} (1980) 2227;
  R.~N.~Mohapatra and G.~Senjanovi\'c,
  ``Neutrino Masses And Mixings In Gauge Models with Spontaneous Parity
  Violation,''
  Phys.\ Rev.\ D {\bf 23} (1981) 165.

\bibitem{adjoint}
  P.~Fileviez~P\'erez,
  ``Renormalizable Adjoint SU(5),''
  Phys.\ Lett.\  B {\bf 654} (2007) 189
  [arXiv:hep-ph/0702287].

\bibitem{TypeIII}
  R.~Foot, H.~Lew, X.~G.~He and G.~C.~Joshi,
  ``Seesaw neutrino masses induced by a triplet of leptons,''
  Z.\ Phys.\ C {\bf 44} (1989) 441.

\bibitem{SUSY-adjoint}
  P.~Fileviez~P\'erez,
  ``Supersymmetric Adjoint SU(5),''
  Phys.\ Rev.\  D {\bf 76}, 071701 (2007)
  [arXiv:0705.3589 [hep-ph]].


\bibitem{CDM}
  M.~Cirelli, N.~Fornengo and A.~Strumia,
  ``Minimal dark matter,''
  Nucl.\ Phys.\  B {\bf 753} (2006) 178
  [arXiv:hep-ph/0512090];
M.~Cirelli, A.~Strumia and M.~Tamburini,
  ``Cosmology and Astrophysics of Minimal Dark Matter,''
  Nucl.\ Phys.\  B {\bf 787} (2007) 152
  [arXiv:0706.4071 [hep-ph]].

\bibitem{test1}
  P.~Fileviez P\'erez,
  ``Fermion mixings vs d = 6 proton decay,''
  Phys.\ Lett.\  B {\bf 595} (2004) 476
  [arXiv:hep-ph/0403286].

\bibitem{WiseCL}
  M.~Claudson, M.~B.~Wise and L.~J.~Hall,
  ``Chiral Lagrangian For Deep Mine Physics,''
  Nucl.\ Phys.\  B {\bf 195} (1982) 297.

\bibitem{Cabibbo}
  N.~Cabibbo, E.~C.~Swallow and R.~Winston,
  ``Semileptonic hyperon decays,''
  Ann.\ Rev.\ Nucl.\ Part.\ Sci.\  {\bf 53} (2003) 39
  [arXiv:hep-ph/0307298].



\bibitem{lattice}
  S.~Aoki {\it et al.}  [JLQCD Collaboration],
  ``Nucleon decay matrix elements from lattice QCD,''
  Phys.\ Rev.\  D {\bf 62} (2000) 014506
  [arXiv:hep-lat/9911026].
See also:
  Y.~Aoki, C.~Dawson, J.~Noaki and A.~Soni,
  ``Proton decay matrix elements with domain-wall fermions,''
  Phys.\ Rev.\  D {\bf 75} (2007) 014507
  [arXiv:hep-lat/0607002].


\bibitem{lowerbounds}
  K.~Kobayashi {\it et al.}  [Super-Kamiokande Collaboration],
  ``Search for nucleon decay via modes favored by supersymmetric grand
  unification models in Super-Kamiokande-I,''
  Phys.\ Rev.\  D {\bf 72} (2005) 052007
  [arXiv:hep-ex/0502026].

\bibitem{Wise}
  A.~V.~Manohar and M.~B.~Wise,
  ``Flavor changing neutral currents, an extended scalar sector, and the  Higgs
  production rate at the LHC,''
  Phys.\ Rev.\  D {\bf 74} (2006) 035009
  [arXiv:hep-ph/0606172];
  M.~I.~Gresham and M.~B.~Wise,
  ``Color Octet Scalar Production at the LHC,''
  Phys.\ Rev.\  D {\bf 76} (2007) 075003
  [arXiv:0706.0909 [hep-ph]].


\bibitem{Pheno-Octets}
  A.~R.~Zerwekh, C.~O.~Dib and R.~Rosenfeld,
  ``A new signature for color octet pseudoscalars at the LHC,''
  arXiv:0802.4303 [hep-ph];
  M.~Gerbush, T.~J.~Khoo, D.~J.~Phalen, A.~Pierce and D.~Tucker-Smith,
  ``Color-octet scalars at the LHC,''
  arXiv:0710.3133 [hep-ph].


\bibitem{Octet-Higgs}
  A.~V.~Manohar and M.~B.~Wise,
  ``Modifications to the properties of a light Higgs boson,''
  Phys.\ Lett.\  B {\bf 636} (2006) 107
  [arXiv:hep-ph/0601212];
  R.~Bonciani, G.~Degrassi and A.~Vicini,
  ``Scalar Particle Contribution to Higgs Production via Gluon Fusion at NLO,''
  JHEP {\bf 0711} (2007) 095
  [arXiv:0709.4227 [hep-ph]].

\bibitem{Frampton}
  P.~H.~Frampton, S.~Nandi and J.~J.~G.~Scanio,
  ``Estimate Of Flavor Number From SU(5) Grand Unification,''
  Phys.\ Lett.\  B {\bf 85} (1979) 225.

\bibitem{Pkalyniak}
  P.~Kalyniak and John N.~Ng
  ``Symmetry Breaking Patterns In Su(5) With Nonminimal Higgs Fields,''
  Phys. \ Rev. \ D {\bf 26}, 890 (1982).

\bibitem{Peckert}
  P.~Eckert, J.~M.~Gerard, H.~Ruegg, T.~Schuecker
 ``Minimization Of The SU(5) Invariant Scalar Potential For The
   Fortyfive-Dimensional Representation,''
  Phys. \ Lett. \ B {\bf 125}, 385 (1983).

\end{thebibliography}
\end{document}